**Characteristics of Magnetic Holes in the Solar Wind Revealed by Parker Solar Probe**

L. Yu[1], S. Y. Huang[1,2], Z. G. Yuan[1], K. Jiang[1], Q. Y. Xiong[1], S. B. Xu[1], Y. Y. Wei[1], J. Zhang[1], and Z. H. Zhang[1]

[1]School of Electronic Information, Wuhan University, Wuhan, 430072, China

[2]Corresponding author: shiyonghuang@whu.edu.cn

**Abstract**

We present a statistical analysis for the characteristics and radial evolution of linear magnetic holes (LMHs) in the solar wind from 0.166 to 0.82 AU using Parker Solar Probe observations of the first two orbits. It is found that the LMHs mainly have a duration less than 25 s and the depth is in the range from 0.25 to 0.7. The durations slightly increase and the depths become slightly deeper with the increasing heliocentric distance. Both the plasma temperature and the density for about 50% of all events inside the holes are higher than the ones surrounding the holes. The average occurrence rate is 8.7 events/day, much higher than that of the previous observations. The occurrence rate of the LMHs has no clear variation with the heliocentric distance (only a slight decreasing trend with the increasing heliocentric distance), and has several enhancements around ~0.525 AU and ~0.775 AU, implying that there may be new locally generated LMHs. All events are segmented into three parts (i.e., 0.27, 0.49 and 0.71 AU) to investigate the geometry evolution of the linear magnetic holes. The results show that the geometry of LMHs are prolonged both across and along the magnetic field direction from the Sun to the Earth, while the scales across the field extend a little faster than along the field. The present study could help us to understand the evolution and formation mechanism of the LMHs in the solar wind.

1. Introduction

Magnetic holes (MHs), which are also called magnetic decreases, are structures with significant decreases/depletions in magnetic field magnitude, usually with durations from several to hundreds of seconds. Linear magnetic holes (LMHs), as a subset of MHs, are defined with no or little change in the magnetic field direction at the boundaries. MHs were first detected in the solar wind by Turner et al. (1977), and then observed in many other interplanetary environments, such as magnetosheath (e.g., Tsurutani et al., 2011; Huang et al., 2017a, 2017b, 2018; Yao et al., 2017), magnetotail plasma sheet (e.g., Sun et al., 2012; Huang et al., 2019), cusp region (e.g., Shi et al., 2009).

Winterhalter et al. (1994) have used the Ulysses data to investigate the relation between the LMHs and the instability of the mirror mode in the range from 1 AU to 5.4 AU to explore the formation mechanism of the holes. They found that the thermal pressure of the plasma inside the holes was generally larger than the surrounding environment, and the ion beta ($\beta$) around the magnetic holes was much higher than the ambient solar wind. Meanwhile, $T_{i\perp}/T_{i\|}$ around the holes was either greater than 1 or larger than the average values of $T_{i\perp}/T_{i\|}$ in the ambient solar wind. Thus, Winterhalter et al. (1994) suggested that the LMHs might be remnants of the mirror mode structures. Stevens and Kasper (2007) have used the Wind data from 1994 to 2004 to perform a scale-free analysis of magnetic holes in the solar wind at 1 AU. Their results showed that the magnetic holes are pressure-balanced structures on all scales. Combined with the measurements of temperature anisotropy, Stevens and Kasper (2007) suggested that the magnetic holes observed at 1 AU are stable remnants of magnetic pressure depletions generated in the source region near the Sun.

Sperveslage et al. (2000) have studied the magnetic holes from 0.3 AU to 17 AU using the data from Helios 1 & 2 and Voyager 2, and found that the LMHs account for 30% of all events, and the surroundings of the holes are in a high β environment. The width of the magnetic holes was reduced from 50 times of the proton inertial length at 0.3~0.4 AU, to 15 times proton inertial length beyond 10 AU. Zhang et al. (2008, 2009) investigated the LMHs in the solar wind at 0.72 AU using Venus Express data, and examined the characteristic size and shape of the LMHs. They defined two parameters, i.e., the width and the eccentricity ($e$) to describe the size and shape, and concluded that the LMHs are two dimensional and more elongated along the field direction, like a rotating ellipse. Zhang et al. (2009) studied the isolated holes and the train of holes later, and shown that the isolated holes are slightly smaller in width and more elongated than multiple holes. Thus, they deduced the evolutionary history of mirror mode structures in which trains of holes merge into isolated structures when moving away from the Sun. Moreover, Xiao et al. (2010) investigated the geometrical structure of LMHs at 1 AU from 2001 to 2004 observed by the Cluster. They found that the geometrical structure of the LMHs is elongated along the magnetic field, and the occurrence rate of LMHs is about 3.7 events/day, which is close to the results (i.e., 4.2 events/day) of Zhang et al. (2008) at 0.72 AU. Based on the point that the occurrence rate and the geometrical shape of interplanetary LMHs have no significant change from 0.72 AU to 1 AU in comparison with Zhang et al.'s work (2009), Xiao et al. (2010) claimed that most of interplanetary LMHs observed at 1 AU are formed and fully developed before 0.72 AU.

However, Russell et al. (2008) have studied the occurrence rate of mirror modes from 0.34 AU to 8.9 AU, and found that the mirror mode occurrence rate linearly diminishes with increasing heliocentric distance $R$ from the Sun, while the durations linearly increase and their relative depth remains constant. Recently, Volwerk et al. (2020) investigated the magnetic characteristics of magnetic holes in the solar wind using the MESSENGER measurements, and found that the occurrence rate of LMHs decreases from Mercury to Venus with the increase of heliocentric distance.

As Parker Solar Probe (PSP) flied unprecedentedly closer to the Sun (0.166 AU), it is an opportunity to investigate the evolution and the characteristics of LMHs from locations near the Sun to the locations far away from the Sun. In the present study, we perform a statistical study on the LMHs using the PSP data in the solar wind from 0.166 to 0.82 AU.

2. **Data Analysis**

The Parker Solar Probe (PSP) mission launched in August, 2018, carrying four suites of instruments. The Solar Wind Electrons, Alphas and Protons (SWEAP) instrument suite (Kasper et al. 2016; Case et al., 2020) is designed to measure velocity distributions of solar wind electrons, alpha particles, and protons. The suite includes the Solar Probe Cup (SPC) and Solar Probe Analyzers (SPANs). The Fields Experiment (FIELDS) instrument suite (Bale et al., 2016), with five voltage probes, two fluxgate magnetometer and one search coil magnetometer, measures electric and magnetic fields from DC to beyond the electron cyclotron frequency.

The magnetic field data from FIELDS and the plasma data from SPC during the first two orbits of the mission (from October, 6$^{th}$, 2018 to December, 19$^{th}$, 2018 and from February, 20$^{th}$, 2019 to May, 15$^{th}$, 2019) are used in this study. The magnetic field data have a resolution of 0.0034 s during the encounter phase and 0.22 s under the cruise mode, while the plasma data have a resolution of 0.873 s for the encounter mode and 27.9 s for the cruise mode. Besides, the data of Ephemeris with the resolution of 1 s are used to obtain the spacecraft location. All vector data, including magnetic field and proton velocity, are presented in Rotational-Tangential-Normal (*RTN*) coordinate system (*R* points away from the sun center toward the spacecraft, *T* is the cross product of solar rotation axis and *R*, and *N* is the cross product of *R* and *T*) in this study.

The criteria for the identification of the MHs used by different researchers are various. Winterhalter et al. (1994) defined MHs with amplitude depression ($B_{min}/B$) less than 0.5, while Zhang et al. (2008) and Xiao et al. (2010) used the criteria that $B_{min}/B$ is smaller than 0.75.

To identify the LMHs, we use both automatic selecting program and manual selection. We take the criteria of $B_{min}/B_{ave} < 0.75$ and the directional change angle $\omega < 15°$, the same as those used by Zhang et al. (2008) and Xiao et al. (2010). The classification of single holes and trains of holes are same to Zhang et al. (2008, 2009): when at least two magnetic holes are found in a 300 s window, they are defined as a train of magnetic holes. However, in this study, we still count a train of magnetic holes as a single event, and represent it by the largest hole in the train.

Firstly, we take the same method as Zhang et al. (2009) and Xiao et al. (2010) to automatically select the events quickly and roughly. We continuously scan the magnetic field data with a slide window of 300 s. Each turn, the center point of the window moves one data point and makes the judgement. If the current point is the minimum in this 300 s interval, we designate it as the local $B_{min}$, and then calculate the average magnetic field $B_{ave0}$ and standard deviation magnetic field $\delta B$ of this interval. The two time points with magnetic field strength closest to $B_{ave0} - \delta B$ at each side of the minimum point are defined as the boundaries of the hole, marked as the start time $t_S$ and the end time $t_E$. Later, the directional change angle $\omega$ is calculated from these two vectors. Events satisfying the criteria $B_{min}/B_{ave0} < 0.75$ and $\omega < 15°$ are selected.

Secondly, in order to identify the LMHs more accurately, we re-determine the start and end time through manual experience and re-select the events (only for those events with relatively steady ambient field background). $t_S$ and $t_E$ are now determined by the start time to decrease and end time to increase of magnetic magnitude of the hole, respectively. The duration of LMHs $\Delta t$ can be obtained. Then, the magnetic field $B_{bef}$ and $B_{aft}$ are determined by time interval $[t_S - 0.5\Delta t, t_S]$ and $[t_E, t_E + 0.5\Delta t]$, respectively. The new rotation angle, $\omega' = <B_{bef}, B_{aft}>$, is used to rejudge the LMHs. Besides, to remove the influence of the decreased holes on the average magnetic field, we take the average field value $B_{ave}$ of the time interval $\{[t_S -150, t_S], [t_E, t_E +150]\}$ to replace $B_{ave0}$. The output parameters of events which satisfy $B_{min}/B_{ave} < 0.75$ and $\omega' < 15°$, include $B_{min}/B_{ave}$, $\omega'$, $B_{min}$, $B_{ave}$, and the time duration $\Delta t$ of the LMHs being crossed by the spacecraft.

## 3. Results

Out of the data of first two orbits we used in this study, we selected 1381 LMH events, with various heliocentric distances from 0.166 AU to 0.82 AU.

Figure 1 shows the typical example of a single hole and a train of magnetic holes. In these magnetic holes, at least one component of the magnetic field decreases significantly. The $\omega$ doesn't change much across the depression region, i.e., less than 15°.

To examine the relationship between the magnetic field magnitude and the heliocentric distance, Figure 2 displays 2-D joint distribution of the event number of the average magnetic field strength $B_{ave}$ as a function of the heliocentric distance $R$. It is clearly seen that the magnetic field magnitude decreases with the increase in $R$, which is consistent with the general prediction that the solar wind magnetic field decreases in strength as it moves away from the Sun, in an approximately $R^{-1}$ dependence.

Figure 3 presents the histograms of the event number of LMHs as a function of the average magnetic field strength $B_{ave}$, the depth of the LMHs $D$ (defined as $1-B_{min}/B_{ave}$), and the durations of the LMHs $\Delta t$, in order to visualize the basic characteristics of the LMH events. We note that the average magnetic field strength is usually not very large (mostly smaller than 16 nT), the depth is concentrated in the range from 0.25 to 0.5, and the magnetic holes mainly have a duration less than 16 s.

The spatial scales of the magnetic holes are analyzed. We use the solar wind velocity to estimate spatial scale by the time duration $\Delta t$ and the average solar wind velocity $V_f$, i.e., $S = V_f \cdot \Delta t$. Furthermore, the scale is normalized by proton gyro-radius $\rho_p$. Here, the proton gyro-radius is calculated according to the average values of the magnetic field and the proton temperature during the time interval $\{[t_S-150, t_S], [t_E, t_E+150]\}$. As aforementioned, the plasma data have a time resolution of 0.873 s for the encounter mode and 27.9 s for the cruise mode. In addition, there are some time intervals when the magnetic data are valid, while the plasma data are not available. Thence, only 985 events with valid plasma data are plotted in Figure 4, which presents the histograms of the spatial scale and the normalized scale of LMHs. One can see that these magnetic holes have a spatial scale mainly in the range of 1778 km to 5623 km, with the peak value of around ~ 3160 km, and the median value of 3164 km. The normalized scale of LMHs mainly distributes between 15 and 70 $\rho_p$, implying LMHs are usually MHD scale structures in the solar wind.

In order to investigate the geometry of magnetic holes, we used the same method as Zhang et al. (2008). We determine the shape of the LMHs by examining their durations and scales as a function of $B_r/B$ where $B_r$ is the radial component of the average magnetic field at the beginning of the hole and $B$ is the average field strength of the magnetic fields before ($B_{bef}$) and after ($B_{aft}$) the holes. The results are shown in Figure 5. Each point represents an individual event, and the diamonds are the median values for each of the 0.1 $B_r/B$ bins. The positive and inverted triangle represent the upper and lower quartiles, respectively. Based on Zhang et al.'s (2008) work, we also assume the shape of the magnetic hole is a rotational ellipsoid, and the cross section of the hole traversed by the satellite is an ellipse, with two characteristic sizes. We take the size along the field and the size across the field, as the long axis and the short axis of the ellipse respectively, then the aspect ratio and the eccentricity of the ellipse can be obtained from two characteristic sizes. The aspect ratio is defined as the ratio of the characteristic size along the field and across the field. The eccentricity is also derived from the characteristic size along the magnetic field (represents the long axis of the ellipse) and across the field (represents the short axis). Assumed that the aspect ratio is $\alpha$, the eccentricity $e$ can be calculated from the following formula: $e = \text{sqrt}(\alpha^2-1)/\alpha$. Following the method of Zhang et al. (2008), we use the best fitted curves on the median values to estimate the asymptotic scales along the field (when $B_r/B = 1$) and across the field (when $B_r/B = 0$), and then obtain the aspect ratio and eccentricity. Note that the eccentricity is calculated only by the aspect ratio of the actual spatial scales $L$.

Figure 5a shows the distribution of the durations as a function of $B_r/B$ for 1381 events, which is fitted by $\Delta t^2 = 136(B_r/B)^2 + 64$. Then, we obtained two characteristic time: the duration along the field is 14.1 s when $B_r/B = 1$, and the duration across the field is 8.0 s when $B_r/B = 0$, yielding that the aspect ratio of the durations is about 1.76. Figure 5b and 5c show the distribution of the spatial scale and the normalized scale of LMHs as a function of $B_r/B$ for 985 events, respectively. From the fitted curve $L^2 = 1.8 \times 10^7 (B_r/B)^2 + 7.3 \times 10^6$ in Figure 5b, we can get that the length along the field is 5030 km, while the length across the field is 2702 km, yielding that the aspect ratio of the spatial scale is 1.86. Thus, the ellipticity of the LMH is 0.84. Using the same method, we got the fitted curve in Figure 5c, namely $Lp^2 = 9037 (B_r/B)^2 + 1806$, and obtained the normalized scale along the field 104 $\rho_p$ and the normalized scale across the field 42 $\rho_p$. Thus, the elongation ratio of the major axis and minor axis of the ellipsoid is 2.48.

We then calculate the plasma parameters both inside and surrounding (or outside) the magnetic holes to investigate the plasma environments. The surrounding region of the hole is defined by the time interval $\{[t_S-150, t_S], [t_E, t_E+150]\}$, while the inside part is defined by the time interval $[t_S, t_E]$. If the durations of magnetic holes are less than the time resolution of the plasma data, it implies that there are no valid plasma data for such magnetic holes. Considering the plasma data cadence and the limited duration of the events as mentioned above, we finally got 490 events with valid data inside of the LMHs. Figure 6 displays the ratios of proton temperature, proton density and proton velocity between inside and outside of the LMHs. One can find that the proton velocity inside the LMHs shows no significant difference comparing with the background plasma, but the proton temperature and density inside the holes are much different with the ones outside of the holes. We also show the results of classified statistics in Table 1 in details. The temperature of 297 events is higher than the one outside of the holes, accounting for 61%. The number of the events with higher density inside the holes is 302, accounting for 62%. 238 events show increase both in temperature and in density, indicating that the increase in temperature and density occurs nearly in half of the holes.

In order to investigate the evolution of the LMHs in the interplanetary space, we analyzed the durations, the depths, the number, and the occurrence rate of the holes in different heliocentric distances.

Figure 7 shows 2-D joint distribution of the event number as a function of the heliocentric distance and the duration (Figure 7a) and the depth (Figure 7b) of the holes. It can be seen that most magnetic holes have a duration between 2 and 25 s, and the duration slightly increase with the increase of the heliocentric distance. The depth has a main range from 0.25 to 0.7, and also become slightly deeper when moving away from the Sun. The LMHs with the depth greater than 0.5 account for 42% of all events.

Figure 8 presents the number and the occurrence rate of the LMHs as a function of the heliocentric distance. One can find that the number of LMHs varies with the increase of the heliocentric distance $R$ without a regular trend, and has an obvious peak around 0.775 AU. Dividing the event number of the LMHs in each bin by the dwell time of PSP observations (Figure 8b) for the corresponding heliocentric distance, one can obtain the occurrence rate of the LMHs in Figure 8c. It seems that the occurrence rate is tend to slightly decrease with the increasing heliocentric distance (not very obvious). The minimum and maximum occurrence rate is 0.16 and 0.60 events/h, respectively, implying that the holes occur 3.8 and 14.4 events/day. The low occurrence rate around ~ 0.175 AU may be caused by the large fluctuations in magnetic field in the inner heliosphere (e.g., Huang et al., 2020) which leads to the difficulty in the identification of the LMHs and the limited number of the LMHs.

Considering that the heliocentric distance and the average magnetic field strength show an inverse relationship, we also show 2-D joint distribution between the average interplanetary magnetic field and the duration (Figure 9a) and depth (Figure 9b) of the LMHs. It is clear seen that the majority of the LMHs are all below 25 s with a wide distribution at low magnetic field strength, but shorten as the increase of the magnetic field strength. However, the depth of the LMHs doesn't have regular variation with the magnetic field strength. The distribution of the depth is broad below 20 nT, and the majority of the LMHs have the depth below 0.6.

## 4. Discussions and Conclusion

Volwerk et al. (2020) has analyzed the magnetic characteristics of magnetic holes in the solar wind between Mercury and Venus, i.e., from 0.3 to 0.7 AU, and showed the 2-D joint distribution of the occurrence rate of the duration (i.e., width) and the depth of the LMHs as a function of the heliocentric distance from the Sun and of the background magnetic field strength. They found that the durations of the LMHs mainly distribute between 5 and 30 s with the median around 10 s, corresponding to ~12-28 proton gyro-radius, and slightly increase with the increase of heliospheric distance, and the depth of the LMHs is spread up to 0.85. There is a broadening of the distribution of the depth of the LMHs as a function of $B_{ave}$ which can be fitted by an exponential function for the boundary of the high occurrence rate. In our results, the LMHs have mainly a duration from 2 and 25 s with the average around 8 s and a main range from 0.25 to 0.7 of the depth. The durations slightly increase and the depth become a little deeper when moving away from the Sun. In addition, the durations have a wide distribution at low magnetic field magnitude $B_{ave}$, and became slightly shorter as the increase of $B_{ave}$, while the depth has no obvious variation with the $B_{ave}$. Thus, one can conclude that the duration of the LMHs in our study have the similar features with the results at 0.3-0.7 AU obtained by Volwerk et al. (2020), while the features of the depth of the LMHs in our study are different with the ones in Volwerk et al. (2020). Several possibilities for these differences: i) the different criteria for the selection of the LMHs in two studies: smaller rotational angle $\omega' \leq 10°$ and higher depth $\delta B/B_{ave}$ >0.5 are used in Volwerk et al. (2020), while rotational angle $\omega' \leq 15°$ and higher depth $\delta B/B_{ave}$ >0.25 are used in this study; ii) different heliocentric distance from the Sun for the statistical study: from 0.3 to 0.7 AU for Volwerk et al.'s study, and from 0.166 to 0.82 AU for this study; iii) different period accompanied with different solar activities which may affect the generation of the LMHs: the data from MESSENGER between 2007 and 2011 are used in Volwerk et al. (2020), and while the PSP's data from October, 6[th], 2018 to December, 19[th], 2018 and from February, 20[th], 2019 to May, 15[th], 2019 are used in this study; iv) more database in Volwerk et al. (2020) than in present study.

Actually, Volwerk et al. (2020) have found a clear slow decrease in the occurrence rate of the LMHs with the increase of the heliocentric distance from 0.3 to 0.7 AU with the average occurrence rate of 2.2 events/day. In this study, the occurrence rate slightly decreases from 0.166 to 0.82 AU, with enhancements around ~0.525-0.575 AU, and ~0.775 AU. Thus, one cannot refer that the LMHs observed before 0.166 AU are formed and fully developed, but it implies that the new LMHs may be locally generated along the way from the Sun to the Earth.

Based on the same criteria for the selection of the LMHs as Zhang et al. (2008, 2009) and Xiao et al. (2010), we compare the occurrence rate, shape and the geometry of the LMHs among these studies. Zhang et al. (2008, 2009) and Xiao et al. (2010) separately analyzed the geometry of LMHs at 0.72 AU and 1 AU. The detailed comparisons between our results with their results are given in Table 2. One can find that the occurrence rate seems to decrease as the heliocentric distance increases, while the shape of the holes doesn't show apparent trait from the near Sun to the near Earth. However, we should point out that these studies used the data at different time, from different missions, and during different solar activities, thus the statistical results may be not suitable to compare with each other.

To further explore the evolution of the geometry of MHs with the heliocentric distance during the flyby of PSP, the total 1381 events (including 985 events with valid plasma data) are divided into three parts by heliocentric distance: the first part of 414 events (including 332 events with valid plasma data) with distance in range of 0.166 to 0.38 AU, the second part of 416 events

(including 296 events with valid plasma data) with distance in range of 0.38 to 0.60 AU, the third part of 551 events (357 events with valid plasma data) with distance ranging from 0.60 to 0.82 AU. Thus, the characteristics of magnetic holes are presented at the mean distances of 0.27 AU, 0.49AU and 0.71 AU, respectively. The results are shown in Table 3. It can be found that the time durations of the holes increase as they move away from the sun, both along and across the ambient magnetic field, as well as the actual spatial scale. As for the normalized spatial scales, the values decrease along the ambient magnetic field, but increase and then decrease across the field from 0.27 AU to 0.71 AU, perhaps due to the change of the local plasma environment (for example, the proton gyro-radius obviously increases with the increase heliocentric distance, not shown here). Moreover, the eccentricity of the ellipsoid corresponding to the holes obviously decreases with the increase heliocentric distance, indicating that the scales across the field increase faster than that along the field.

In summary, we studied the characteristics of the LMHs, including the duration, the spatial scale, the geometry, and the evolution with the change of heliocentric distance. Our results could help us to understand the evolution and formation mechanism of the LMHs in the solar wind.

Based on the above analysis, the results are summarized as follows:
1) More LMHs occur at low interplanetary magnetic field strength. The time durations are shortened as the increase of interplanetary magnetic field strength.
2) The average occurrence rate of the LMHs is about 8.66 events/day, higher than previous studies. The occurrence rate has a slightly decrease with the increasing heliocentric distance but with several peaks around ~0.525 AU and 0.775 AU, implying that the LMHs may also be locally generated after 0.166 AU.
3) The plasma temperature and density inside the LMHs are higher than the ambient ones for > 60% of the LMHs, while the velocity has no obvious change inside and outside of the LMHs. About 50% of the holes show increase both in proton temperature and in density.
4) The durations of the LMHs mainly distribute between 2 and 25 s with the average around 8 s, and slightly increase with the increase of the heliocentric distance. The depths mainly vary from 0.25 to 0.7, and become slightly deeper when moving away from the Sun.

5) As the LMHs travel away from the Sun, the durations and the spatial scales increase both along and across the magnetic field, and the eccentricity of the ellipsoid corresponding to the holes decreases with the increase of heliocentric distance.

**Acknowledgement**

This work was supported by the National Natural Science Foundation of China (41674161, 41874191, 41925018) and the National Youth Talent Support Program. We thank the entire PSP team and instrument leads for data access and support. The SWEAP and FIELDS investigation and this publication are supported by the PSP mission under NASA contract NNN06AA01C. PSP data is publicly available from the NASA's Space Physics Data Facility (SPDF) at https://spdf.gsfc.nasa.gov/pub/data/psp/.

**Table 1.** The results of classified statistics for the temperature and density of the inside and outside plasma environment.

| Category1 | Category2 | Amount (490 in total) | Rate |
|---|---|---|---|
| $T_{pin} / T_{pout} > 1$ | $N_{pin} / N_{pout} > 1$ | 238 | 49% |
| | $N_{pin} / N_{pout} < 1$ | 59 | 12 % |
| $T_{pin} / T_{pout} < 1$ | $N_{pin} / N_{pout} > 1$ | 64 | 13% |
| | $N_{pin} / N_{pout} < 1$ | 129 | 26% |

**Table 2.** Comparison of the durations and spatial scales among Zhang et al. (2008), Xiao et al. (2010) and our work. The eccentricity is calculated by the ratio of the spatial scale *L*. "/" represents no data.

| Study | Our Work | Zhang et al. (2008) | Xiao et al. (2010) |
|---|---|---|---|
| Spacecraft | PSP | Venus Express | Cluster-C1 |
| Distance | 0.166-0.82 AU | 0.72 AU | 1 AU |
| *Dt* (along) | 14.1 s | 13.08 s | 10.1 s |
| *D*t (across) | 8.0 s | 5.39 s | 5.5 s |
| Ratio (*Dt*) | 1.76 | 2.43 | 1.84 |
| *L* (along) | 5030 km | / | 5329 km |
| *L* (across) | 2702 km | / | 2757 km |
| Ratio (*L*) | 1.86 | / | 1.93 |
| *Lp* (along) | 104 $\rho_p$ | 107 $\rho_p$ | / |
| *Lp* (across) | 42 $\rho_p$ | 42 $\rho_p$ | / |
| Ratio (*Lp*) | 2.48 | 2.55 | / |
| Eccentricity | 0.84 | / | 0.86 |
| Event Number | 1381 | 791 | 567 |
| Occurrence Rate | 8.7 events/day | 4.2 events/day | 3.7 events/day |

**Table 3.** Comparison of the durations and spatial scales of the three segments. The eccentricity is calculated by the ratio of the spatial scale $L$.

| Heliocentric distance | 0.16~0.38 AU | 0.38~0.60 AU | 0.5~0.82 AU |
|---|---|---|---|
| Average Distance | 0.27 AU | 0.49 AU | 0.71 AU |
| $Dt$ (along) | 11.8 s | 16.1 s | 17.7 s |
| $Dt$ (across) | 4.0 s | 8.3 s | 9.4 s |
| Ratio ($Dt$) | 2.95 | 1.94 | 1.88 |
| $L$ (along) | 3563 km | 6798 km | 5447 km |
| $L$ (across) | 1445 km | 3287 km | 3321 km |
| Ratio ($L$) | 2.47 | 2.07 | 1.64 |
| $Lp$ (along) | 136 $\rho_p$ | 134 $\rho_p$ | 60 $\rho_p$ |
| $Lp$ (across) | 37 $\rho_p$ | 42 $\rho_p$ | 41 $\rho_p$ |
| Ratio ($Lp$) | 3.68 | 3.19 | 1.46 |
| Eccentricity | 0.91 | 0.88 | 0.79 |

**Figure Captions**

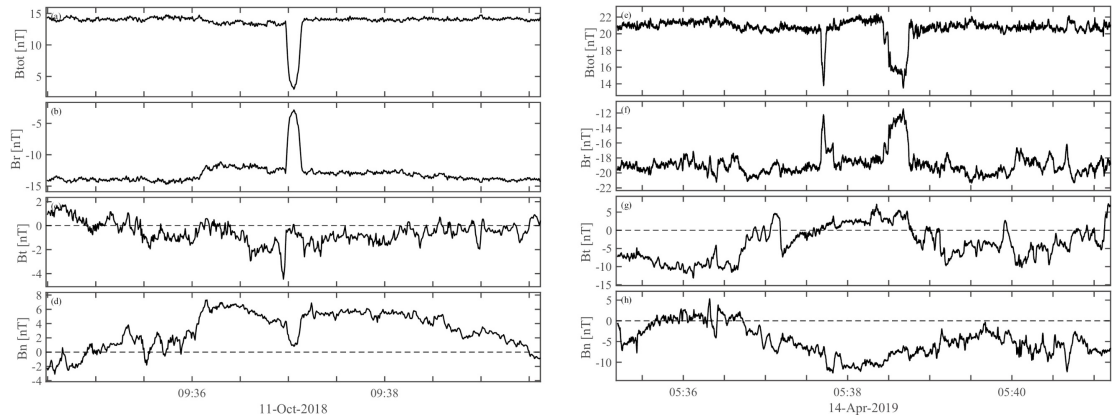

Figure 1. Examples of a single linear magnetic hole (LMH) and a train of LMHs in RTN coordinates.

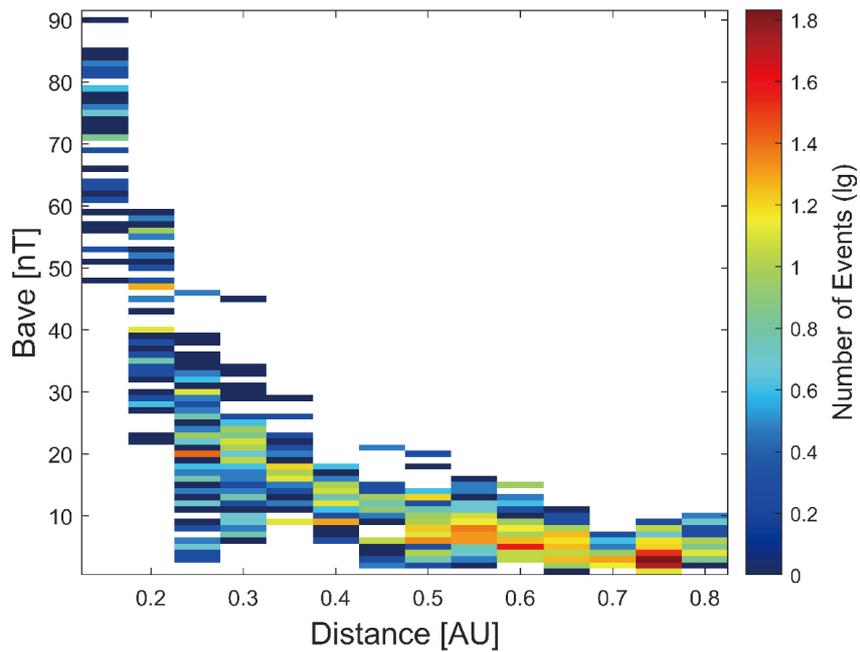

Figure 2. Joint distributions of the average magnetic field strength $B_{ave}$ as a function of heliocentric distance $R$ for all LMH events. The bins are 0.05 AU in $R$ and 1 nT for $B_{ave}$.

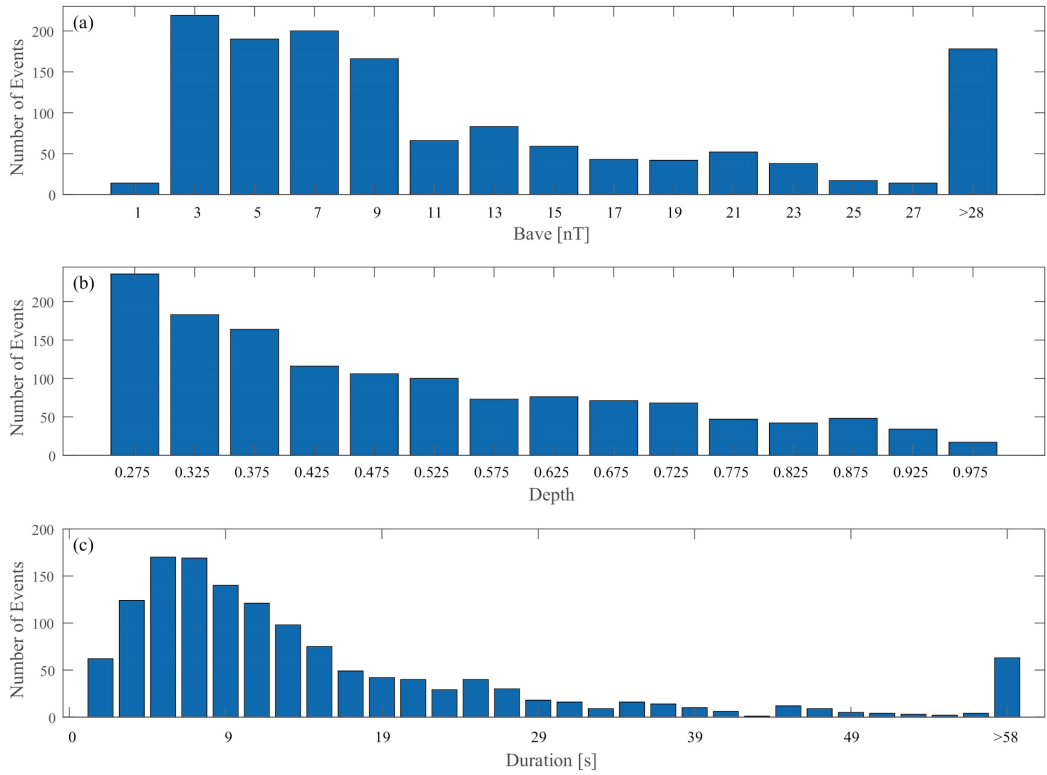

Figure 3. Histograms of the event number as a function of the average magnetic field strength, the depth, and the durations of the LMHs.

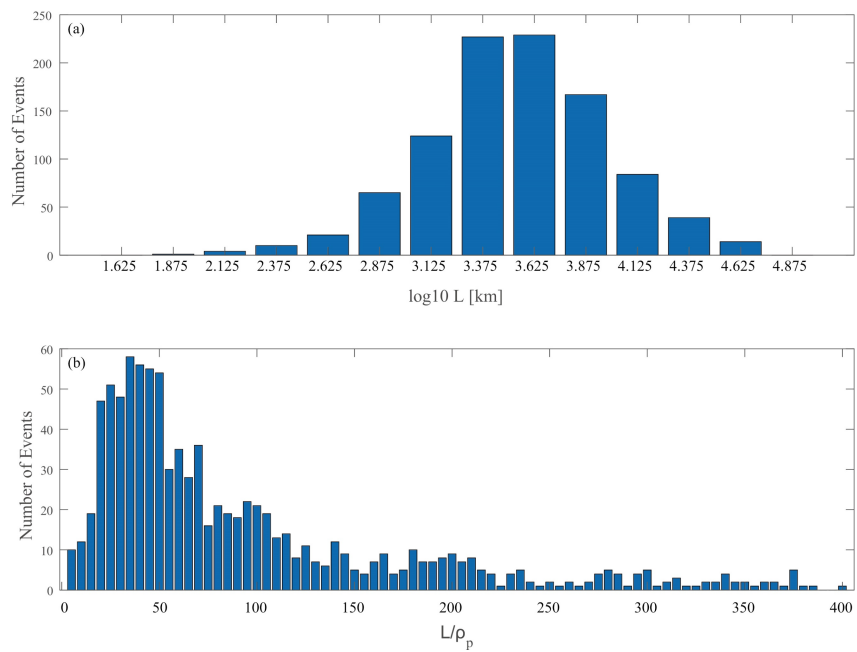

Figure 4. Histograms of the event number as a function of the spatial scale and the normalized scale of the LMHs.

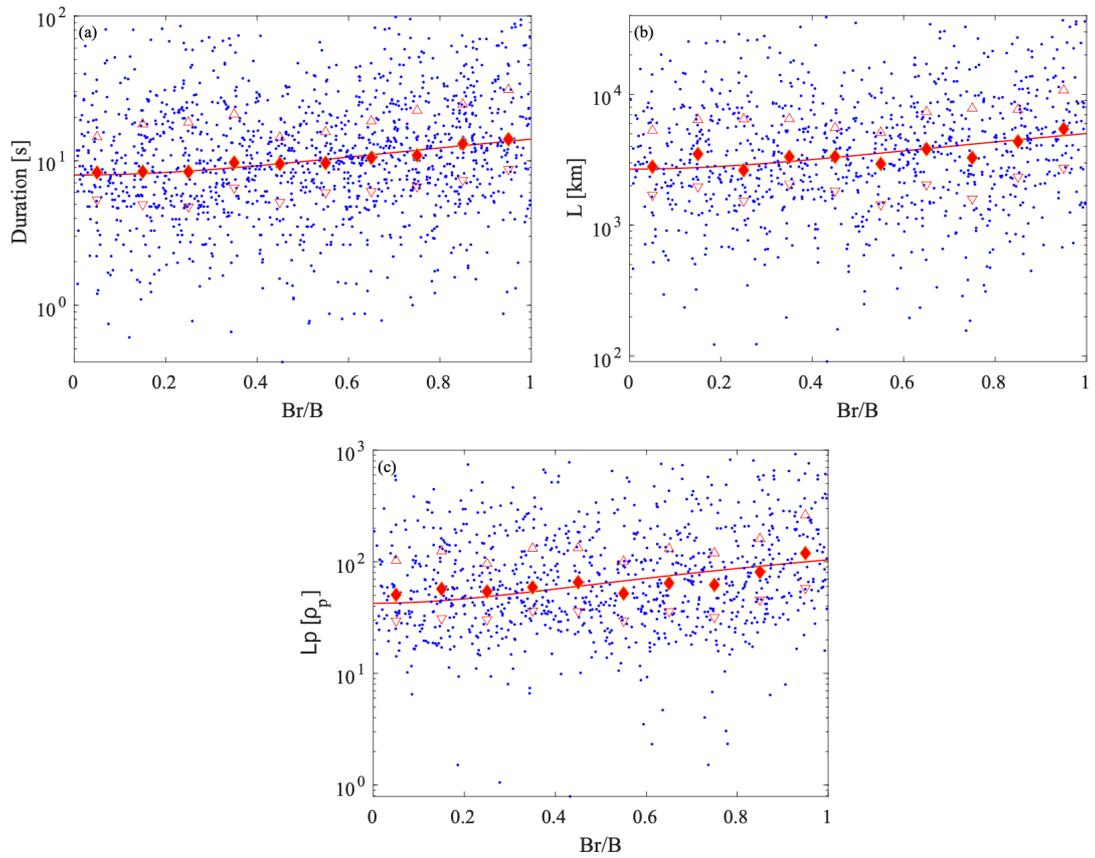

Figure 5. The distributions of (a) the durations, (b) the spatial scale, and (c) the normalized scale of LMHs as a function of $B_r/B$. Each point represents an individual event and diamonds are the median values for each of the 0.1 $B_r/B$ bins. The positive and inverted triangle represent the upper and lower quartiles, respectively. The curves are the best fitting results to the median points.

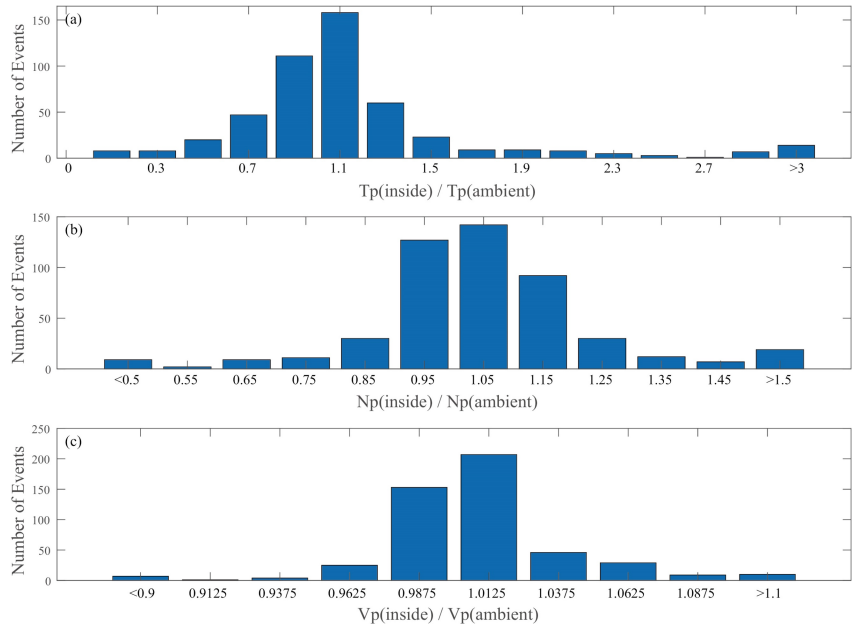

Figure 6. Histograms of the event number as a function of the ratio of plasma parameters (including proton temperature, density and velocity) between inside the LMHs and the surrounding environments (i.e. outside of the LMHs).

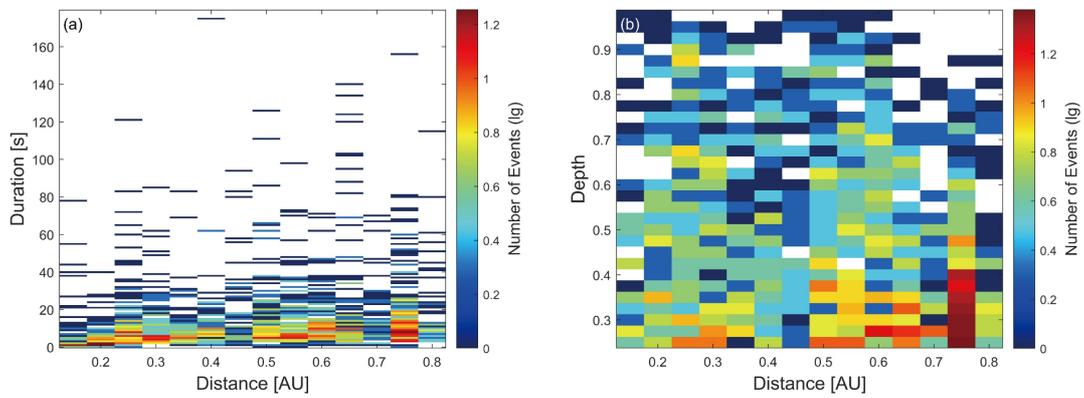

Figure 7. Joint distributions of the event number between the heliocentric distance and (a) the durations and (b) the depth of the LMHs.

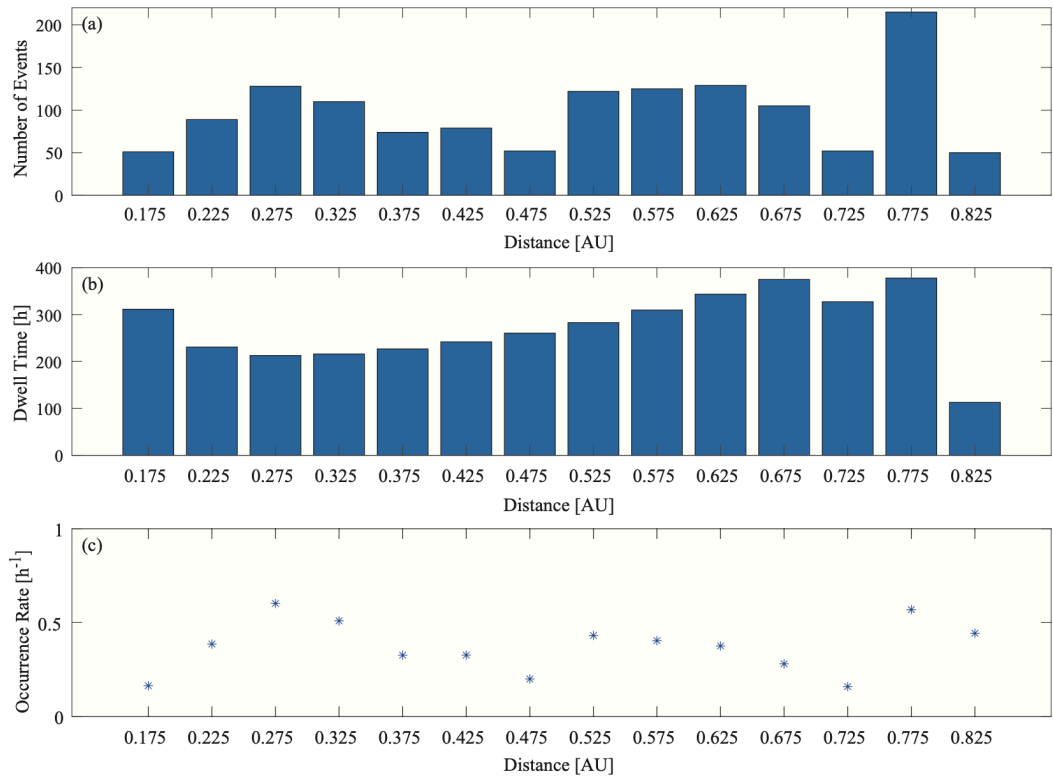

Figure 8. (a) Histogram of the event number of the LMHs as a function of the heliocentric distance; (b) the dwell time of PSP along the heliocentric distance; (c) the occurrence of the LMHs as a function of the heliocentric distance.

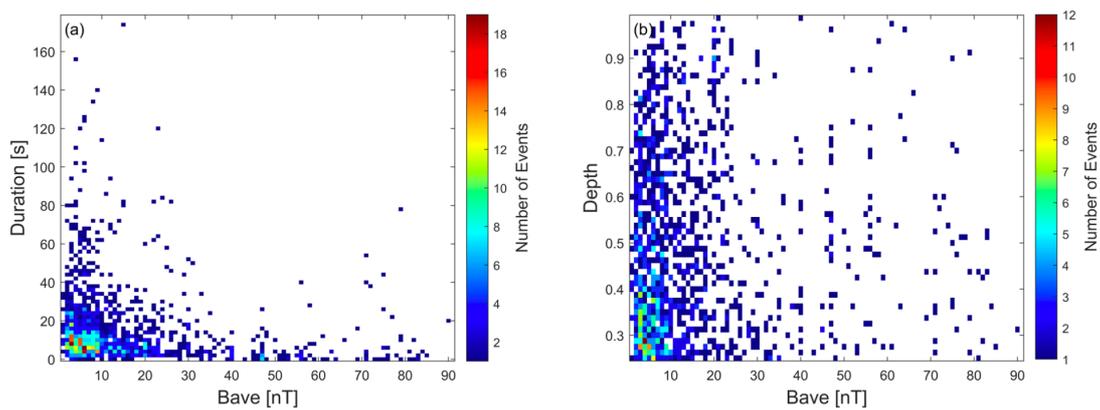

Figure 9. Joint distributions of the event number between the average magnetic field $B_{ave}$ and (a) the durations of the LMHs and (b) the depth of LMHs.